\documentclass{article}

\usepackage[preprint]{neurips_2022}

\usepackage[utf8]{inputenc} 
\usepackage[T1]{fontenc}    
\usepackage{hyperref}       
\usepackage{url}            
\usepackage{booktabs}       
\usepackage{amsfonts}       
\usepackage{nicefrac}       
\usepackage{microtype}      
\usepackage{xcolor}         

\usepackage{enumitem}
\makeatletter
\newcommand{\printfnsymbol}[1]{%
  \textsuperscript{\@fnsymbol{#1}}%
}
\usepackage{algorithmicx,algpseudocode,algorithm}
\usepackage{wrapfig,tikz}
\usepackage{amsmath,longtable,fancyhdr,booktabs,multirow,graphicx,float}
\usepackage{adjustbox, bigstrut, tabularx, multirow, makecell, diagbox}
\usepackage{amssymb,xcolor,amsthm}
\usepackage{subcaption}

\newcommand{\be}{\begin{equation}}
	\newcommand{\ee}{\end{equation}}

\definecolor{Gray}{gray}{0.85}
\definecolor{LightCyan}{rgb}{0.88,1,1}

\newcolumntype{a}{>{\columncolor{Gray}}c}

\makeatletter
\usepackage{xspace}
\def\@onedot{\ifx\@let@token.\else.\null\fi\xspace}
\DeclareRobustCommand\onedot{\futurelet\@let@token\@onedot}

\title{Diagnosis Assistant for Liver Cancer Utilizing a Large Language Model with Three Types of Knowledge}

\author{%
    Xuzhou Wu\\
    SIGS, Tsinghua University \\ 
    \texttt{wuxz21@mails.tsinghua.edu.cn} 
    \And
    Guangxin Li\\
    Radiotherapy Department,Beijing Tsinghua Changgung Hospital \\ 
    \texttt{liguangxin2006@163.com} 
    \AND
    Xing Wang  \\ 
    Radiotherapy Department, Beijing Tsinghua Changgung Hospital \\ 
    \texttt{wangxing0901@163.com}
    \AND
    Zeyu Xu  \\ 
    Radiotherapy Department, Beijing Tsinghua Changgung Hospital \\ 
    \texttt{13801272617@163.com}
    \AND
    Yingni Wang  \\ 
    School of Biomedical Engineering, Tsinghua University \\ 
    \texttt{wangyn23@mails.tsinghua.edu.cn}
    \AND
    Jianming Xian  \\ 
    SIGS, Tsinghua University \\ 
    \texttt{19121757854@163.com}
    \AND
    Xueyu Wang  \\ 
    SIGS, Tsinghua University \\ 
    \texttt{wangxuey23@mails.tsinghua.edu.cn}
    \AND
    Gong Li  \\ 
    Radiotherapy Department, Beijing Tsinghua Changgung Hospital \\ 
    \texttt{dr\_gongli@163.com}
    \And
    Kehong Yuan\thanks{Corresponding author.}  \\ 
    SIGS, Tsinghua University \\ 
    \texttt{yuankh@sz.tsinghua.edu.cn} 
}

\begin{document}
\maketitle
\begin{abstract}
    \noindent {\bf Background:}      
    Liver cancer has a high incidence rate, but there is a lack of experienced liver cancer doctors in primary healthcare settings. Advances in large models and other artificial intelligence technologies offer the potential to assist less experienced doctors in diagnosing liver cancer.\\
    {\bf Purpose:} 
    To overcome some of the current limitations of large models in the field of liver cancer diagnosis, such as inadequate understanding of medical images for specific diseases, insufficient consideration of liver blood vessels, and the inability to ensure the accuracy of the medical information used by large models. The goal is to provide a specialized diagnostic assistant for liver cancer, thereby improving the diagnostic capabilities of less experienced doctors in primary healthcare settings.\\
    {\bf Methods:} 
    A liver cancer diagnosis framework combining large models and small models is proposed. Optimized small models are used for more precise perception of patient images. Specifically, for liver tumor segmentation, a segmentation network is proposed that iteratively removes ambiguous pixels with edge enhancement. For liver vessel segmentation, a multi-scale, multi-level differential network is introduced for segmenting small objects. After obtaining this information, features are extracted from the segmentation results and combined with medical records to form the patient's personalized knowledge base. In terms of large model diagnosis, Chain of Thought (COT) technology is used to design prompts that mimic the thinking patterns of experienced doctors in liver cancer diagnosis and treatment. Additionally, Retrieval-Augmented Generation (RAG) technology is employed to provide answers based on reliable domain diagnostic knowledge and trusted past cases from doctors.\\
    {\bf Results:}
  For the small model component of the structure, the proposed liver tumor segmentation and liver vessel segmentation methods show performance improvements over control methods, resulting in more accurate information extraction. For the large model component of the structure,the responses from our proposed liver cancer diagnosis assistant scored over 1 point higher on a 10-point scale in two evaluation metrics, as rated by doctors, compared to control methods.\\
    {\bf Conclusions:} 
    First, our method, by combining two segmentation small models, can more accurately perceive the semantic information of medical images. The two segmentation models improve the classification of ambiguous pixels and the perception of small objects.
    Second, our method has been specifically optimized for the task of liver cancer diagnosis and treatment, such as considering the positions of blood vessels to cater to specific treatment needs.
    Third, our method improves the credibility and interpretability of responses by mimicking the thought processes of experienced doctors and using reliable resources. This approach has been recognized by doctors and is beneficial for the task of liver cancer auxiliary diagnosis.
\\
\end{abstract}

\section{INTRODUCTION}
Liver malignant tumors (liver cancer) are a serious threat to human health, ranking 6th globally in incidence among malignant tumors. However, during the treatment of liver cancer, there is a significant disparity in the distribution of medical resources across regions. Many remote areas lack experienced senior doctors, and the existing medical resources cannot meet the high demand for diagnosis and treatment. With the rapid development of artificial intelligence technologies, including large models, in recent years, this study aims to use such technologies to provide diagnostic suggestions and assist less experienced doctors in primary care settings.

In recent years, large models have made significant advancements in information analysis and processing. For example, GPT-3\cite{brown2020language} pushed the parameter count of models to unprecedented levels, and the subsequent release of ChatGPT had an even greater impact. In the Chinese-speaking domain, models like ChatGLM\cite{zeng2022glm} have also achieved outstanding performance. While general-purpose large language models have achieved remarkable success, diagnostic large models in the medical field have also rapidly progressed. For instance, the Med-Palm model \cite{singhal2022large} has demonstrated the ability to answer medical questions at the level of the United States Medical Licensing Examination (USMLE). Other large models in the medical domain have also been released, including ChiMed-GPT\cite{tian2023chimed} and HuatuoGPT\cite{chen2023huatuogpt}. The Chinese diagnostic medical model MedLLM\cite{bao2023disc} has undergone alignment training to human preferences and values , and has performed excellently in question-answering tasks. Visual models or multimodal models include ChatCAD\cite{wang2023chatcad} and others.

However, while these models possess general capabilities across various diseases, their abilities are not specifically optimized for particular diseases such as liver cancer discussed in this study. Currently, large models excel mainly in understanding textual information or two-dimensional images and some medical images. However, for certain specialized medical images that require high-level knowledge, large models do not sufficiently understand their complex medical semantic information. For example, when addressing specific needs such as considering the protection of liver vessels in combined radiotherapy and surgical treatment of liver cancer, these models lack targeted optimization. Additionally, the accuracy and credibility of the medical materials these models are trained on may be questionable. Different doctors have varying levels of acceptance of these models, and they prefer to use reliable medical information that they have vetted.

Therefore, this study aims to mimic human doctors by diagnosing in a step-by-step manner. First, liver tumors and critical at-risk organs (liver vessels) are segmented. Then, in a manner akin to how doctors perceive and extract features from these image data, the model is guided to focus on this crucial information. These data, along with the medical record information, are then fed into the large model to obtain diagnostic recommendations.

In terms of liver tumor segmentation, the methods for liver tumor segmentation include traditional methods, machine learning, and deep learning methods. Traditional methods include thresholding \cite{choudhary2008entropy, moghe2011automatic}, region growing based on similarity \cite{anter2013automatic}, and semi-automatic segmentation using statistical properties \cite{yuan2017hierarchical, li2012new}. Machine learning methods include adaptive learning methods \cite{shimizu2008ensemble}. Deep learning methods for liver tumor segmentation include fully convolutional network methods \cite{sun2017liver}, cascaded methods for liver and liver tumor segmentation \cite{budak2020cascaded}, methods integrating deformable convolution and spatial pyramid modules \cite{lei2021defed}, multi-scale information extraction module methods \cite{kushnure2021ms}, methods using dense connections \cite{li2017joint}, and attention mechanism methods under different architectures \cite{li2020attention, valanarasu2021medical}.
However, in the process of liver tumor segmentation, there still exists the problem of fuzzy edges leading to poor classification of edge pixels. We introduce the concept of edge enhancement and the idea of iteratively removing ambiguous pixels in fuzzy target recognition, proposing our segmentation network.

In terms of liver vessel segmentation, algorithms can be broadly categorized into traditional methods, machine learning methods, and deep learning methods. Traditional methods often involve user interaction with the segmentation algorithm, such as region growing \cite{oliveira2011segmentation}, Hessian methods \cite{foruzan2012hessian, li2019vessel}, threshold methods \cite{wilson1997segmentation, guo19983}, morphological tracking algorithms \cite{friman2010multiple, guo2020novel}, and K-means algorithms \cite{zeng2018automatic}. Deep learning methods include approaches using dense connections \cite{kitrungrotsakul2017robust}, 3D residual networks \cite{yu2019liver}, methods that enhance semantic links with surrounding tissues using dilated convolutions \cite{xu2020training}, methods to reduce noise from low-quality dataset annotations \cite{xu2021noisy}, and methods using designed attention mechanisms to focus on vessels \cite{yan2020attention}. However, the issue of the thin and difficult-to-extract features of liver vessels has not been specifically addressed, so this study proposes our liver vessel segmentation method.

This study proposes a framework for assisting in the diagnosis of liver cancer, which combines large models and small models to mimic the step-by-step diagnostic process of human doctors. First, small models optimized for specific tasks are used to enhance the understanding of complex medical semantics, segmenting liver tumors and critical at-risk organs (liver vessels). Subsequently, in a manner akin to how doctors perceive and extract features from these image data, the model is guided to focus on this crucial information rather than directly analyzing the entire medical image with a large model. Additionally, during the process, the treatment approach combining radiotherapy and surgery for liver cancer includes consideration of liver vessels, which are typically not accounted for in standard treatments.

This medical image information, along with medical record information, serve as a personalized knowledge base for the patient. To ensure accurate diagnosis by the large model, other two types of knowledge are incorporated: filtered, reliable domain knowledge, and private diagnostic case knowledge from doctors. During the process where the large model combines these three types of knowledge to provide answers, RAG (Retrieval-Augmented Generation) technology is utilized to enhance credibility by answering based on the given knowledge. Meanwhile, a prompt incorporating CoT (Chain-of-Thought) technology, specifically designed for the liver cancer domain, is used to simulate the thinking process of experienced liver cancer doctors, ensuring high-quality responses.
\section{MATERIALS AND METHODS}
\subsection{Data source}

We selected both public and private datasets.
The public dataset is the 3DIRCADB dataset, which includes CT data of 20 liver cases, containing annotations for the liver, liver vessels, and liver tumors.The private dataset comes from the Department of Radiation Oncology at Tsinghua Changgung Hospital, including CT data of 10 liver cases, with annotations for the liver and liver tumors.After slicing these data, they were used for liver vessel segmentation and liver tumor segmentation. The 3DIRCADB dataset, used for the tumor segmentation task, contains 625 images, while the Tsinghua Changgung Hospital dataset contains 197 images. The 3DIRCADB dataset, used for the vessel segmentation task, contains 1688 images. For different target areas of liver tumors annotated by the Radiation Oncology Department, we used the GTV region for our labels. We first trained the liver segmentation model of CENet on abdominal CT slices. After obtaining the region of interest, we then segmented liver tumors and liver vessels within the liver region.

The CT used by the Department of Radiation Oncology at Tsinghua Changgung Hospital is the LightSpeed RT. The slice thickness is 5.0mm, the peak voltage (kVp) is 120kV, the spacing between slices is 5.0mm, and the data collection diameter is 500.0mm.

Medical record data were also collected from Tsinghua Changgung Hospital. The main items included in the medical records are:past treatments, including previous surgical treatment, previous radiotherapy, previous radiofrequency treatment, previous interventional treatment, and previous drug treatment.Patient symptoms, mainly considering abdominal pain, abdominal distension, ascites, jaundice, lower back pain, fever, abdominal mass, etc. Some biochemical indicators of tumor markers, such as AFP and PIVKA, etc. The past treatment cases are also the treatment plans for liver cancer patients at ChangGung Hospital.

\subsection{Overall framework}
For large models, large language models have powerful performance in analyzing medical corpora but only accept text input. Current vision models or multimodal models, although capable of understanding some medical images, do not understand certain images of specific diseases with complex semantic information deeply enough, and they lack rich and reliable knowledge in that domain. Therefore, we propose a framework that combines small models and large language models to provide auxiliary diagnosis for liver cancer.
\begin{figure}[htb!]
	\centering
	\includegraphics[width=0.95\linewidth]{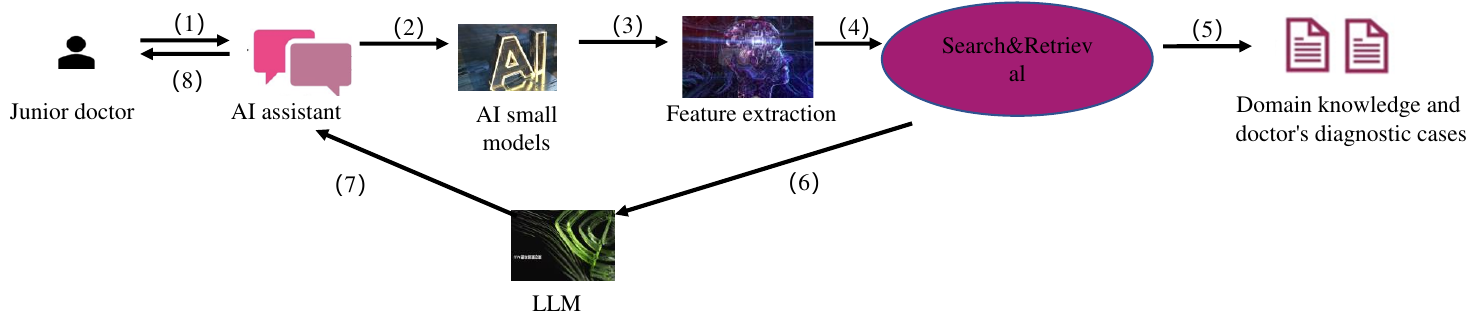}
	\caption{Overall Framework of the Liver Cancer Diagnosis Assistant}
	\label{fig:Figure1}
	
\end{figure}
As shown in Figure~\ref{fig:Figure1} , when in Step 1 the junior doctor inputs the patient’s imaging information and medical record information into the AI assistant, first in Step 2, the AI small model performing the segmentation task is used to segment two important parts in the patient’s CT liver scan, namely liver tumors and liver vessels. Because the segmentation model has been optimized, the segmentation results obtained will be better than those of general segmentation models. Then, in Step 3, imitating the perception of a senior doctor, features of the liver tumors and liver vessels are extracted, specifically the features related to pathological characteristics that doctors focus on when reading images, and these are described in text. Through this step, the patient's imaging information is converted into text information that the large language model can process. The description of the imaging information is combined with the medical record information and jointly input into the next step.Afterwards, to suppress the hallucination phenomenon of the large language model, we utilize RAG technology (retrieval-augmented generation) based on reliable domain knowledge and past diagnostic cases screened by doctors, and feed it into Step 6. The large language model answers based on this content. In the processing by the large language model, we imitate the thinking mode of high-level doctors, design a chain of thought for liver cancer diagnosis and treatment, form an intelligent diagnosis assistant for liver cancer, and return the analysis results to the AI assistant and the junior doctor.

\subsection{Liver tumor segmentation}
During the liver tumor segmentation process, even in enhanced CT images, the tumor areas sometimes have low contrast with the surrounding areas and blurred boundaries, which poses significant challenges for accurate segmentation. Mei et al. proposed a method of gradual refinement iteration in the camouflaged object detection network \cite{mei2021camouflaged}. This study adopts the idea from that method, introduces and modifies the concept of camouflaged object segmentation into the medical image segmentation field, and achieves the goal through step-by-step refined segmentation: feature extraction for detection, blur recognition of targets, and refinement to eliminate false positives and false negatives.

\begin{figure}[htb!]
	\centering
	\includegraphics[width=0.6\linewidth]{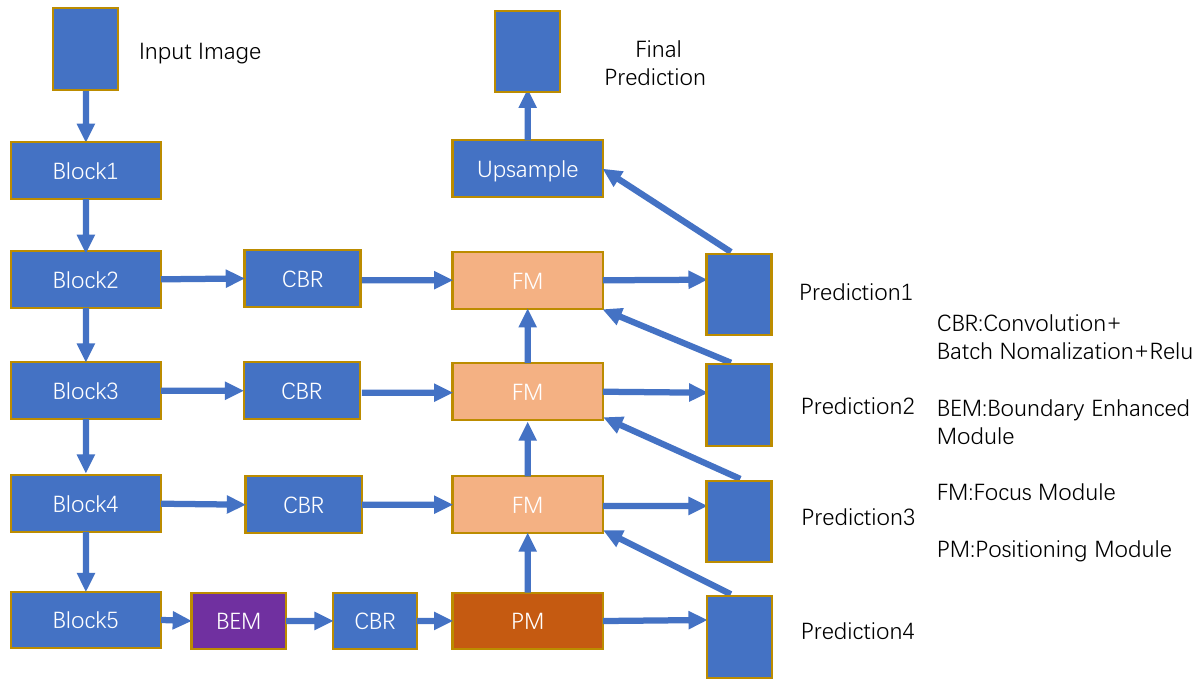}
	\caption{The overall framework of tumor segmentation method}
	\label{fig:Figure2}
\end{figure}

During the liver tumor segmentation process, even in enhanced CT images, the tumor areas sometimes have low contrast with the surrounding areas and blurred boundaries, which poses significant challenges for accurate segmentation. Mei et al. proposed a method of gradual refinement iteration in the camouflaged object detection network \cite{mei2021camouflaged}. This study adopts the idea from that method, introduces and modifies the concept of camouflaged object segmentation into the medical image segmentation field, and achieves the goal through step-by-step refined segmentation: feature extraction for detection, blur recognition of targets, and refinement to eliminate false positives and false negatives,as shown in Figure~\ref{fig:Figure2}

\begin{figure}[htb!]
	\centering
	\includegraphics[width=0.6\linewidth]{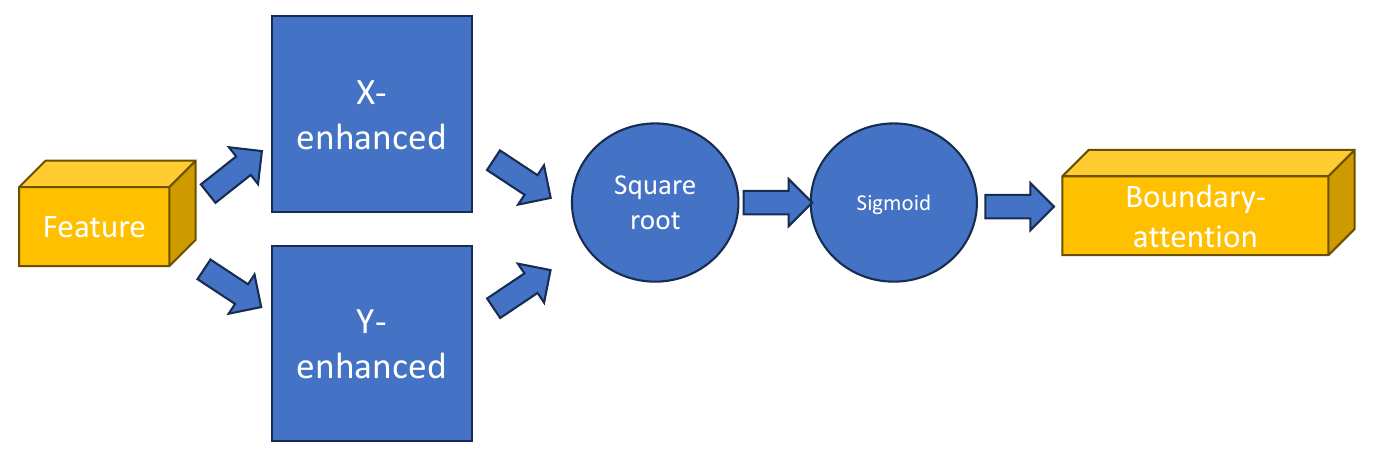}
	\caption{BEM Module}
	\label{fig:Figure3}
\end{figure}

In this network, given the characteristic of blurry edges in liver tumors, the use of the Sobel operator's concept to enhance edges before initial positioning can make the location results more accurate, as shown in Figure~\ref{fig:Figure3}. The specific principle of the Boundary Enhanced Module (BEM) is that the weights of the convolution layers are set as the filtering operators of the Sobel filter, enhancing the image edges in the horizontal and vertical directions, respectively. These convolution layers do not set bias terms. Then, the square root of the horizontal and vertical gradients is normalized to obtain pixel-level enhanced edge weights, which are multiplied pixel-by-pixel with the feature map, allowing subsequent networks to focus more on the structural information in the image, guiding the segmentation. The operator values in the image can be represented by the following formula.

\begin{equation}
	X_{-} \text{enhanced} = \left[\begin{array}{ccc}
		-1 & 0 & 1 \\
		-2 & 0 & 2 \\
		-1 & 0 & 1
	\end{array}\right]
	\label{eq:x边缘}
\end{equation}

\begin{equation}
	Y_{-} \text{enhanced} = \left[\begin{array}{ccc}
		-1 & -2 & -1 \\
		0 & 0 & 0 \\
		1 & 2 & 1
	\end{array}\right]
	\label{eq:y边缘}
\end{equation}

Where $X_{-} \text{enhanced}$ and $Y_{-} \text{enhanced}$ represent the horizontal enhancement operator and vertical enhancement operator, respectively.

\begin{figure}[htb!]
	\centering
	\includegraphics[width=0.6\linewidth]{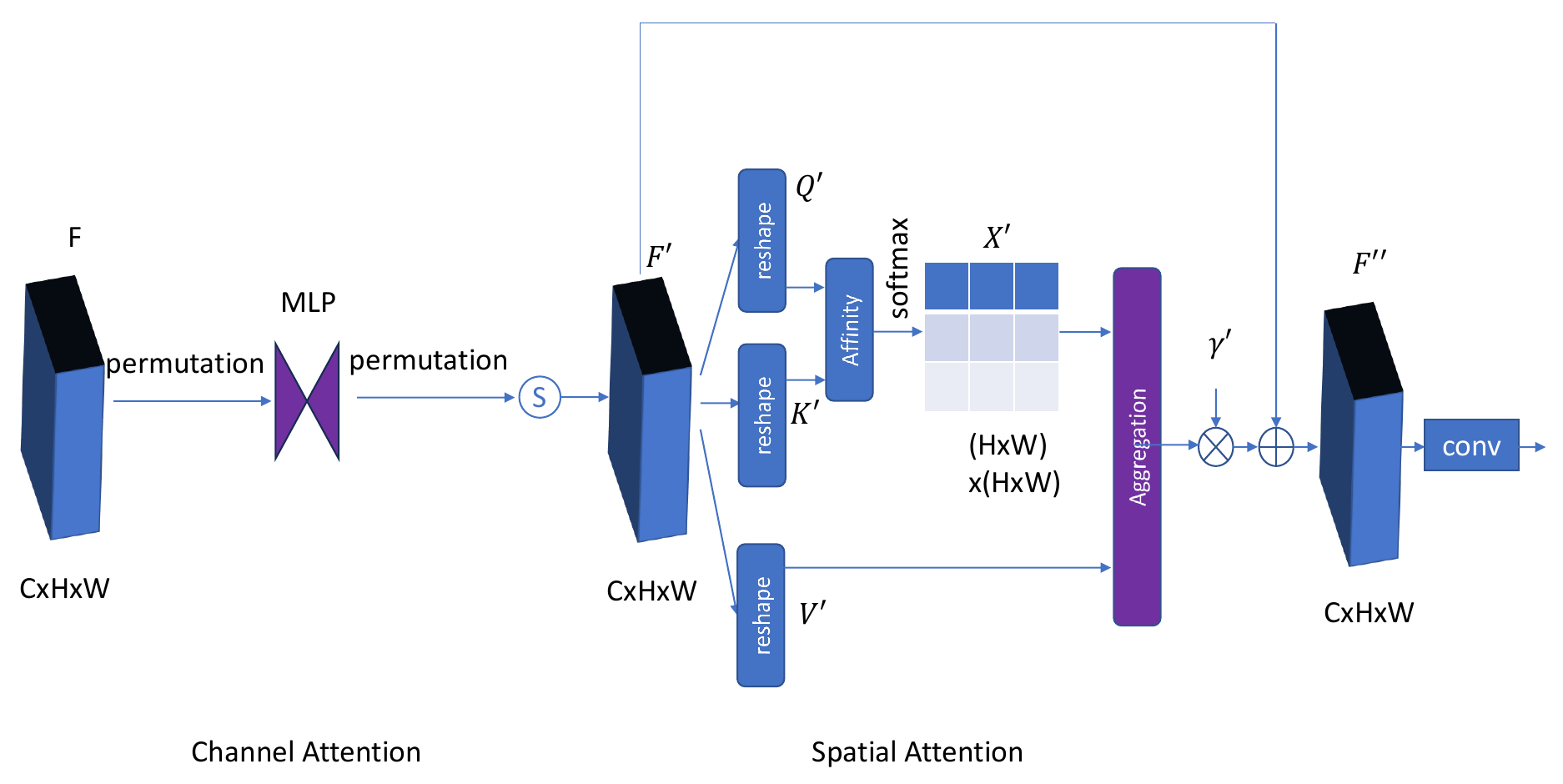}
	\caption{Positioning Module}
	\label{fig:Figure4}
\end{figure}

As shown in Figure~\ref{fig:Figure4}. The Positioning Module (PM) consists of two attention modules: channel attention module and spatial attention module, aimed at capturing long-term relationships in channel and spatial positioning from a global perspective, enhancing high-level semantic features, and obtaining preliminary segmentation results.

Initially, the process passes through a section that computes channel attention, which utilizes the principle of channel attention computation found in the design of GAM attention as described in Liu et al.\cite{liu2021global}. The input features are first subjected to a dimensional transformation, moving the channel dimension to the end. The resulting matrix then passes through a multilayer perceptron (MLP) to reduce the number of channels, followed by another MLP that increases the channels, enhancing the important information within the channels. Finally, the matrix's channel dimension is transformed back to its original position and processed through a Sigmoid activation layer to obtain the final channel attention feature map.

Feed the obtained features into the subsequent spatial attention module, which aims to determine the positions in space that require increased weighting. First, utilize a 1x1 convolution to aggregate and adjust the features in $F_{i:}$. Then, further reshape the features to obtain three features $Q'$, $K'$, and $V'$."

To obtain the spatial attention map, multiply the transpose of $Q'$ with $K'$, and normalize the result. The calculation of the spatial attention $X'$ can be represented as follows:

\begin{equation}
	x_{i j}^{\prime}=\frac{\exp \left(Q_{: i}^{\prime} \cdot K_{: j}^{\prime}\right)}{\sum_{j=1}^N \exp \left(Q_{: i}^{\prime} \cdot K_{: j}^{\prime}\right)}
	\label{eq:2-6空间注意力图}
\end{equation}

Where $Q_{: i}^{\prime}$ represents the i-th column of the matrix $Q'$, and $x_{i j}^{\prime}$ represents the influence of the j-th position on the i-th position. Multiply $V'$ with the transposed matrix $X'$, adjust the dimensions, and obtain the weight map. In the processing of the attention map, a skip connection is added to obtain the output features, as shown in equation.
\begin{equation}
	F_{: i}^{\prime \prime}=\gamma^{\prime} \sum_{j=1}^N\left(V_{: j}^{\prime} x_{j i}^{\prime}\right)+F_{: i}^{\prime}
	\label{eq:2-7skip connection}
\end{equation}
Where $F_{: i}^{\prime \prime}$ represents the obtained output features, $\gamma^{\prime}$ represents learnable parameters. After obtaining the output features, they are passed through a convolutional layer with a size of 7x7 and the result is fed into the FM module.

\begin{figure}[htb!]
	\centering
	\includegraphics[width=0.6\linewidth]{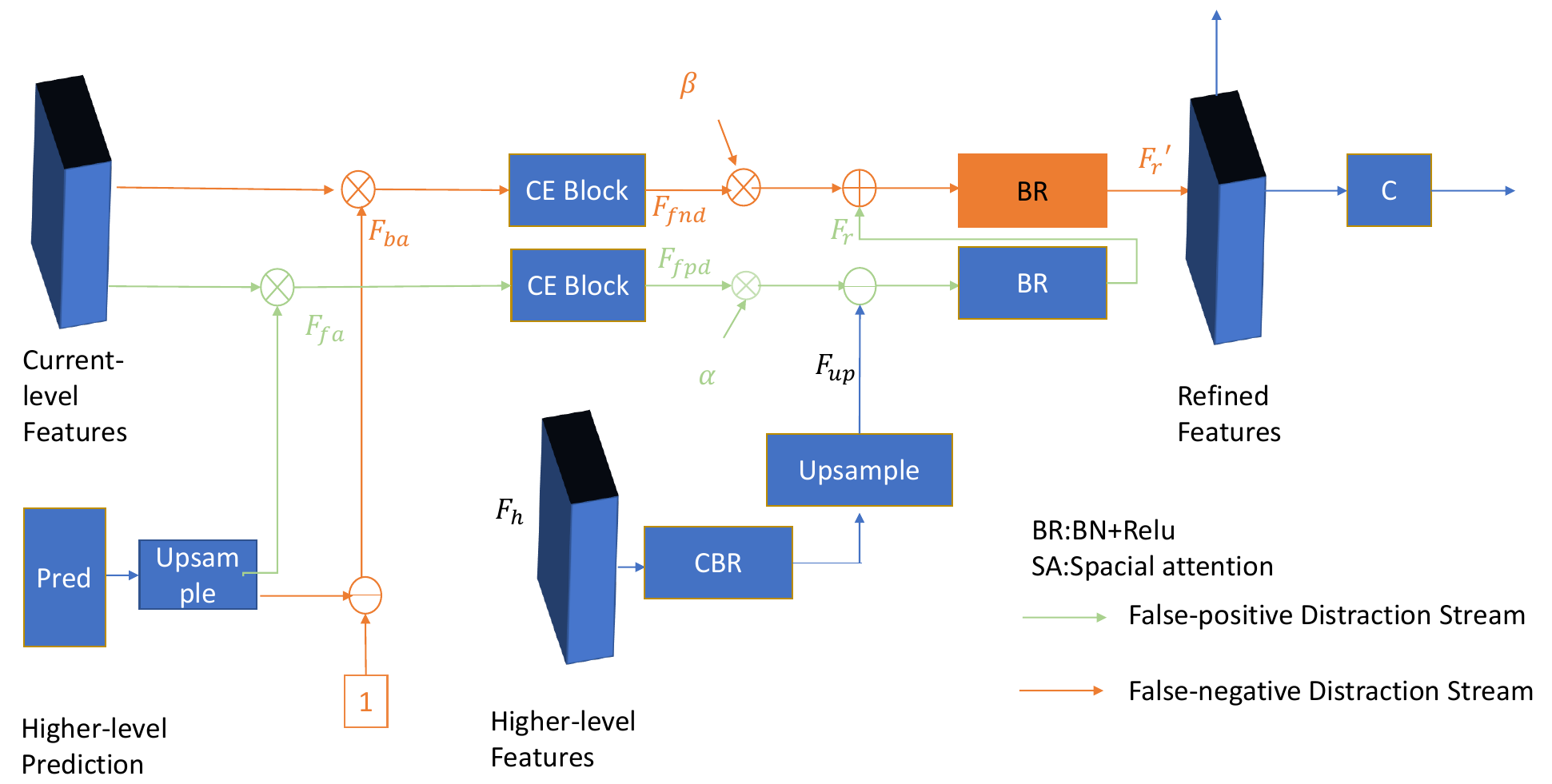}
	\caption{Focusing Module}
	\label{fig: Figure5}
\end{figure}

The design of the focus module is the same as that of PFNET\cite{mei2021camouflaged},which is a module derived from hidden object recognition task. We introduce the methods from object recognition into the field of medical image segmentation to optimize pixel classification errors. As shown in the Figure~\ref{fig: Figure5}, the information from higher-level predictions allows the two branches to focus on background features and foreground features respectively. Through subsequent iterations of the CE BLOCK (context exploration block) and higher-level prediction features, many false positive and false negative pixels in the foreground and background can be eliminated. This aims to reduce the misclassification of pixels caused by the similarity between tumors and the background.

\subsection{Liver vessel segmentation}
\begin{figure}[htb!]
	\centering
	\includegraphics[width=1\linewidth]{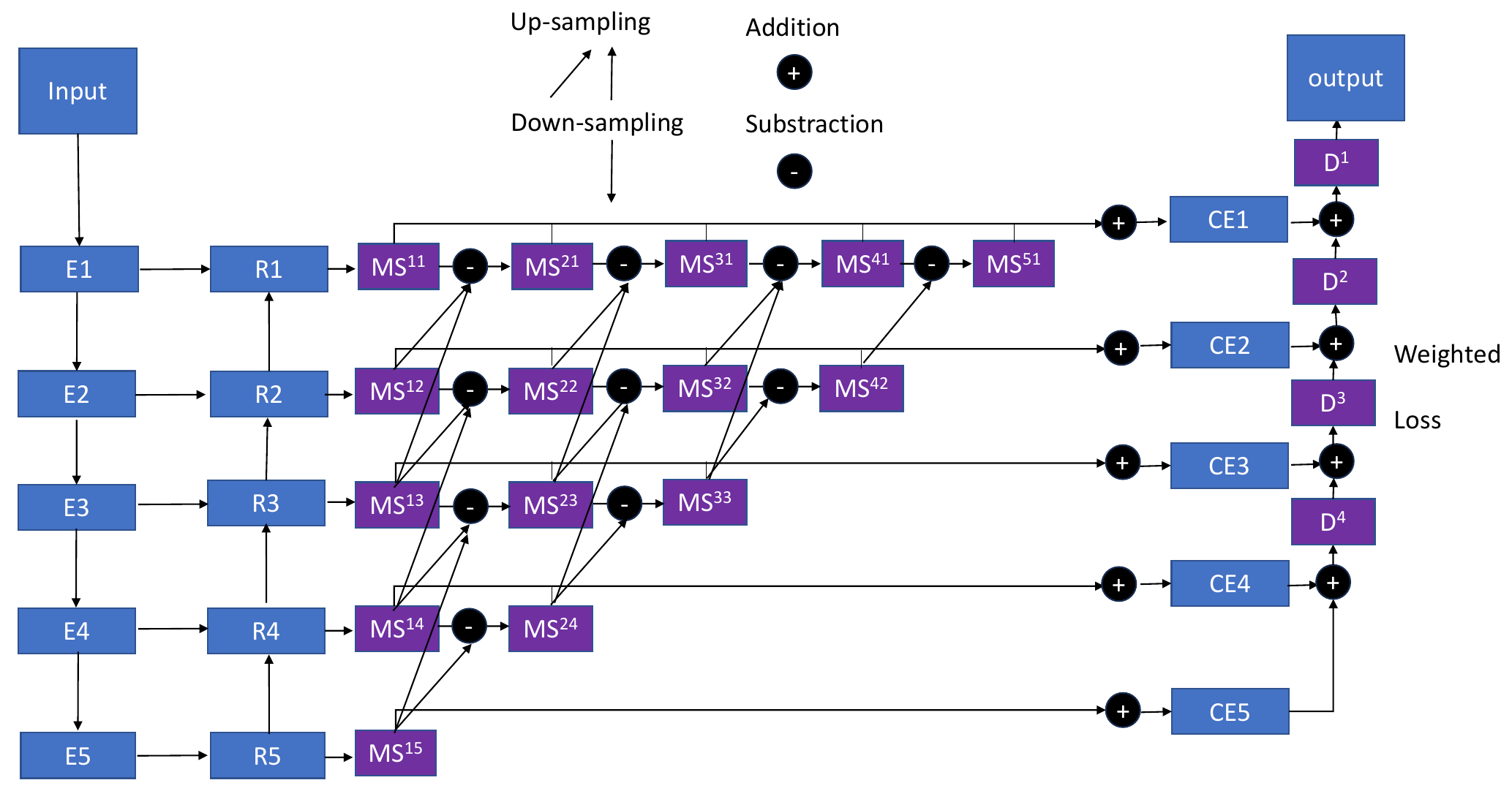}
	\caption{(a)MSMLNet Structure (b)Weighted Structual Loss}
	\label{fig:Figure6}
\end{figure}

The radial size of liver vessels varies greatly, and extracting information from vessels of different thickness may yield poor results. Additionally, simple addition or concatenation of information from different levels in UNET-like architectures can lead to information redundancy and the potential neglect of important details.The M2SNET network has made significant contributions in addressing the issue of redundant information fusion. M2SNET utilizes differential modules with different receptive fields for subtraction-based fusion of features between adjacent levels, focusing on truly important information. Building upon this approach, we further perform similar operations between non-adjacent levels, achieving a multi-scale multi-level differential segmentation network(MSMLNet),as shown in Figure~\ref{fig:Figure6}.

Apart from the information fusion challenge, another issue arises due to the predominantly small size of liver vessel shapes. Information obtained after multiple downsampling layers tends to ignore many small vessels that originally existed, resulting in significant distortion. Conventional methods of information extraction often fail to capture these small vessels, thus our proposed method incorporates improvements specifically targeting this issue.

\begin{figure}[htb!]
	\centering
	\includegraphics[width=0.5\linewidth]{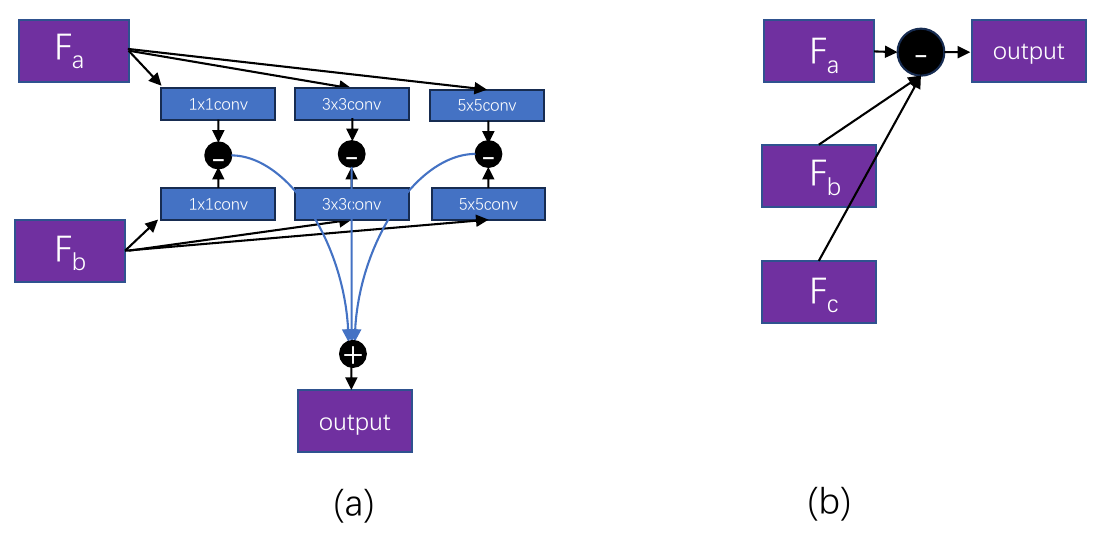}
	\caption{(a)multi-scale differential principle (b)multi-layer differential principle}
	\label{fig:Figure7}
\end{figure}

As shown in Figure~\ref{fig:Figure7},the process of multi-scale differential information extraction can be represented as follows:
\begin{equation}
	Sub_{ab}=\sum_{i \in{1,3,5}}\left|c_i\left(F_a\right)-c_i\left(up\left(F_b\right)\right)\right|
	\label{eq:multi-scale-difference}
\end{equation}
Here, $Sub_{ab}$ represents the differential result between adjacent layers, $F_a$ and $F_b$ represent the feature maps of adjacent layers in the encoder, $up$ denotes upsampling, and $c_i$ represents a convolutional kernel of size $i \times i$.
The process of joint differential extraction across multiple levels can be represented as follows:
\begin{equation}
	Sub_{abc}=Sub_{ab}+0.1 \cdot Sub_{ac}
	\label{eq:multi-level-difference}
\end{equation}
Here, $Sub_{ab}$ represents the differential result between adjacent layers, $Sub_{ac}$ represents the differential result between two layers with a level difference of 2, and $Sub_{abc}$ represents the final differential result.
The design of connecting multiple levels enables non-adjacent levels to fuse information at an early stage, allowing the model to capture global features.

The differences between the loss function and the M2SNET network lie in the varying weights of the structural loss function, which captures the feature differences between predictions and ground truth at different levels using a pre-trained VGG network.The specific formulation of the overall weighted structural loss is as follows:

\begin{equation}
	Loss_{structure}=\frac{64 \times Loss_s^{1}+16 \times Loss_s^{2}+4 \times Loss_s^{3}+Loss_s^{4}}{85}
	\label{eq:structural-loss}
\end{equation}

Here, $Loss_s^{1}$, $Loss_s^{2}$, $Loss_s^{3}$, and $Loss_s^{4}$ represent the structural losses computed for the first, second, third, and fourth layers, respectively. Additionally, one difference between this method and M2SNET is that it does not randomly adjust the image size during data augmentation, instead, it maintains a fixed size of \( 512 \times 512 \).

\subsection{Image feature extraction}
\subsubsection{Tumor diameter}
Extracting tumor diameter information is necessary because it can reflect the category of the tumor's current progression, serving as an important metric that influences the choice of treatment methods.

\subsubsection{Irregularity of the tumor margin}
Irregularity of the tumor margin represents the aggressiveness of the tumor. The largest cross-sectional area of the tumor is selected as the representative for calculating the margin irregularity. The centroid of the lesion is selected, and the distances from the points on the lesion's boundary to the centroid are calculated. Then, the degree of variation in these distances is calculated, as shown in formula x. The method for calculating the irregularity of the tumor margin can be expressed as:
\begin{equation}
	\operatorname{diff}=\frac{1}{(N-1) \cdot d_{\text{avg}}} \sum_{k=1}^{N-1}|d(k+1)-d(k)|
	\label{eq:Irregularity of the tumor margin}
\end{equation}

Where diff represents the irregularity of the margin, N represents the total number of points, k represents the k-th point, d represents the distance from a point to the centroid, and $d_{avg}$ represents the average distance to the centroid.The calculation results will be compared with a threshold to obtain a description of whether the margin irregularity is high or low.

\subsubsection{The contrast in grayscale between the tumor region and the surrounding area}

The contrast in grayscale between the tumor region and the surrounding area can reflect the differences in HU values and blood supply between the tumor region and the surrounding tissues. The slice with the largest cross-sectional tumor area is still selected for the calculation. This metric is divided into two parts for measurement, with the outer boundary representing the sequence of outer boundary points adjacent to the boundary line.

The calculation process for the difference in grayscale between the internal tumor region and the outer boundary region of the tumor can be expressed as:"
\begin{equation}
	g_{\text{differ} 1}=\left|\frac{g_{\text{outboundary}} - g_{\text{in}}}{g_{\text{in}}}\right|
	\label{eq:灰度对比度1}
\end{equation}

Where $\ {g}{\text {differ } 1}$ represents the internal to outer boundary difference coefficient, $\ g{outboundary}$ is the average grayscale value of the outer boundary sequence, and $\ g_{in}$ is the average grayscale value of the internal tumor region.

The grayscale difference between the tumor region and the surrounding region is:
\begin{equation}
	g_{\text{differ} 2}=\left|\frac{g_{\text{outarea}} - g_{\text{in}}}{g_{\text{in}}}\right|
	\label{eq:灰度对比度1}
\end{equation}

Where $\ {g}{\text {differ } 2}$ represents the difference coefficient between the internal tumor region and the surrounding area, $\ g{outarea}$ is the average grayscale value of the surrounding area, and $\ g_{in}$ is the average grayscale value of the internal tumor region.

These two metrics will be compared with a threshold to obtain a description of the grayscale contrast

\subsubsection{Grayscale features of the tumor interior}

The grayscale features of the tumor interior reflect the condition of necrotic areas within the tumor. By calculating the entropy, standard deviation, uniformity, and third-order moment of the grayscale distribution within the tumor, a description of the tumor's internal grayscale features can be obtained.

\subsubsection{Distance between tumor and main blood vessels}

As previously mentioned, when designing a combined radiotherapy and surgical plan, the liver blood vessels are considered  organs at risk to prevent excessive weakening of the vessel walls during radiotherapy, which could affect suturing in subsequent surgeries. Therefore, the distance between the tumor target area and the main blood vessels of the liver needs to be considered. In each cross-section of the liver blood vessels, the largest connected domain is selected and combined to create the main blood vessel. The shortest distance between the tumor and this main blood vessel is then calculated.

\subsection{LLM processing}
The ChatGLM-6B model is utilized, which is based on the General Language Model (GLM) architecture and is a product jointly developed by Tsinghua University and Zhipu AI. This model was chosen because the study focuses on liver cancer treatment in China, with both the input questions and the reference knowledge being in Chinese. And this model has been optimized for Chinese.Besides, its 6B size facilitates deployment in primary hospitals.

To enhance the accuracy of the knowledge learned by the large model, RAG technology is used to make the model forget its existing knowledge and instead respond based on a new reference knowledge base. The knowledge base referenced by the large model includes the CSCO Guidelines for the Diagnosis and Treatment of Primary Liver Cancer (2022), knowledge graphs extracted from some reliable corpora, and processed diagnostic plans for liver cancer patients provided by senior doctors.

The principle of RAG and LangChain is to use the encoder of the large model to vectorize the knowledge base and user questions. Then, the similarity between the question vectors and the knowledge base vectors is compared, and the top k most relevant vectors are returned. These contents serve as references for the model's responses and are also provided to the user to see the reference sources.

\begin{figure}[htb!]
	\centering
	\includegraphics[width=1\linewidth]{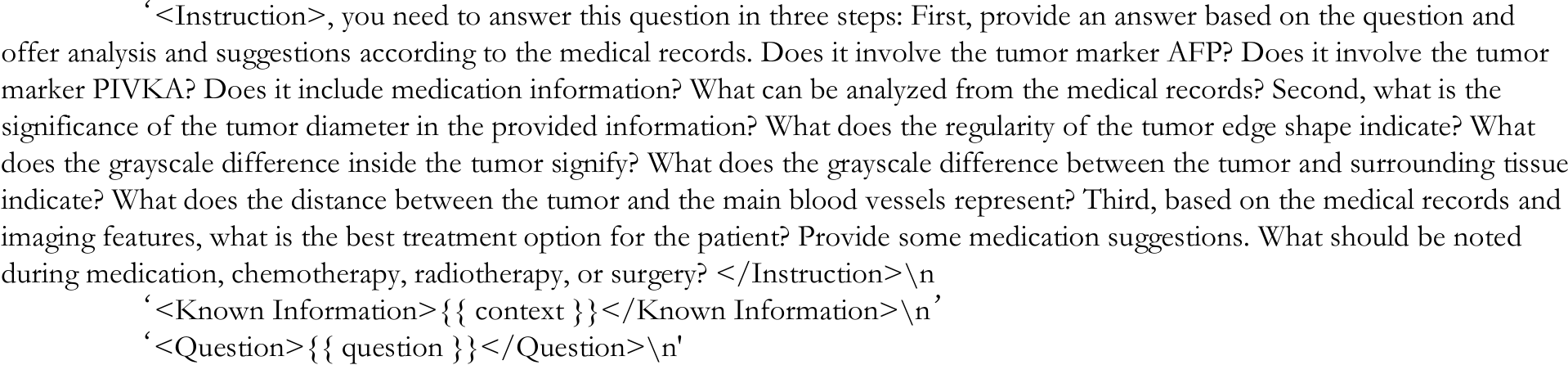}
	\caption{Prompt for liver cancer diagnosis}
	\label{fig:Figure8}
\end{figure}

Using chain of thought(COT) technology allows the model to think step-by-step. The conclusions drawn from this process are more accurate than directly outputting the conclusion, and the thought process behind the output is more interpretable and credible.
The Figure~\ref{fig:Figure8}. shows the prompt designed using chain of thought technology, based on the thought process of experienced liver cancer doctors.

\section{RESULTS}
\subsection{Liver tumor segmentation results}
\begin{figure}[h]
	\centering
	\includegraphics[width=0.8\linewidth]{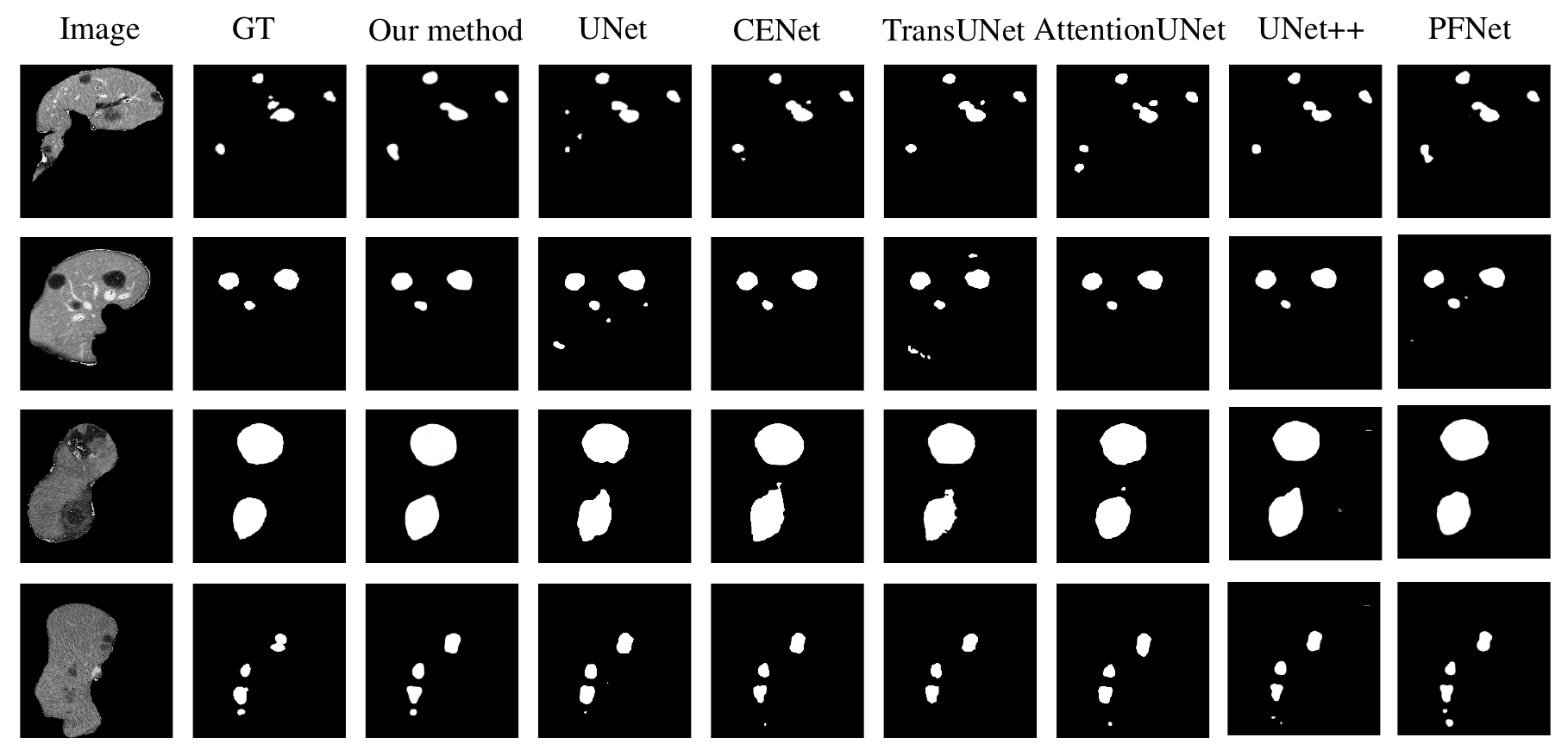}
	\caption{The visualization results of liver tumor segmentation}
	\label{fig:Figure9}
\end{figure}

In the diagnostic process of the liver cancer diagnostic assistant, we involved the segmentation of liver tumors. This study was validated on both a public dataset (3Dircadb dataset) and a private dataset (Tsinghua Changgung Hospital Radiotherapy Department dataset). The initial learning rate was set to 0.001, with dynamic decay, using the Adam optimizer for 400 epochs. The obtained metrics are shown in the Table~\ref{tab:liver tumor seg performance}. Additionally, ablation experiments were conducted, as shown in the Table~\ref{tab:Liver tumor segmentation ablation experiment}. We also performed a visual comparison of our method with other methods, as shown in Figure~\ref{fig:Figure9}.
\begin{table}[ht]
	\centering
	\caption{The tumor segmentation performance of different models on two datasets(A represents the 3Dircadb dataset, and B represents the Tsinghua Chang Geng Hospital Radiation Oncology Department dataset.)}
	\label{tab:liver tumor seg performance}
	\begin{tabular}{c|cc|cc|cc|cc|cc}
		\toprule
		\multirow{2}{*}{Model} & \multicolumn{2}{c|}{DICE} & \multicolumn{2}{c|}{IOU} & \multicolumn{2}{c|}{ACC} & \multicolumn{2}{c|}{REC} & \multicolumn{2}{c}{PRE} \\
		& A & B & A &B & A & B & A & B & A & B \\
		\midrule
		UNet                  & 0.834     & 0.746     & 0.754     & 0.643     & 0.985     & 0.965     & 0.826     & 0.692     & 0.872     & 0.863     \\
		CENet                 & 0.904     & 0.842     & 0.845     & 0.755     & 0.995     & 0.976     & 0.896     & 0.857     & 0.927     & 0.866     \\
		TransUNet\cite{chen2021transunet} & 0.879     & 0.773     & 0.816     & 0.656     & 0.992     & 0.963     & 0.867     & 0.799     & 0.913     & 0.797     \\
		AttentionUNet         & 0.833     & 0.767     & 0.756     & 0.672     & 0.987     & 0.971     & 0.811     & 0.742     & 0.894     & 0.841     \\
		UNet++\cite{zhou2018unet++}       & 0.831     & 0.787     & 0.751     & 0.687     & 0.987     & 0.970     & 0.803     & 0.807     & 0.906     & 0.825     \\
		PFNet                  & 0.903     & 0.855     & 0.846     & 0.767     & 0.994     & 0.975     & 0.908     & 0.883     & 0.909     & 0.856     \\
		Our Method                  & \textbf{0.912}     & \textbf{0.876}     & \textbf{0.853}     & \textbf{0.796}     & \textbf{0.995}     & \textbf{0.978}     & \textbf{0.929}     & \textbf{0.923}     & \textbf{0.904}     & \textbf{0.856}     \\
		\bottomrule
	\end{tabular}
\end{table}

\begin{table}[htb!]
	\centering
	\caption{Liver tumor segmentation ablation experiment}
	\begin{tabular}{cccccc}
		\toprule
		Model        & DICE      & IOU  & ACC    & REC    & PRE          \\
		\midrule
		Our method	&\textbf{0.912}	&\textbf{0.853}	&\textbf{0.995} &\textbf{0.929} &\textbf{0.904} \\
		Our method(without BEM Module)	&0.907	&0.846	&0.993 &0.919 &0.911 \\
		Our method(without PM Module)	&0.898	&0.840	&0.994 &0.901 &0.904 \\
		Our method(without FM Module)	&0.902	&0.841	&0.993 &0.916 &0.898\\
		\bottomrule
	\end{tabular}
	\label{tab:Liver tumor segmentation ablation experiment}
\end{table}

\subsection{Liver vessel segmentation results}
\begin{figure}[htb!]
	\centering
	\includegraphics[width=0.8\linewidth]{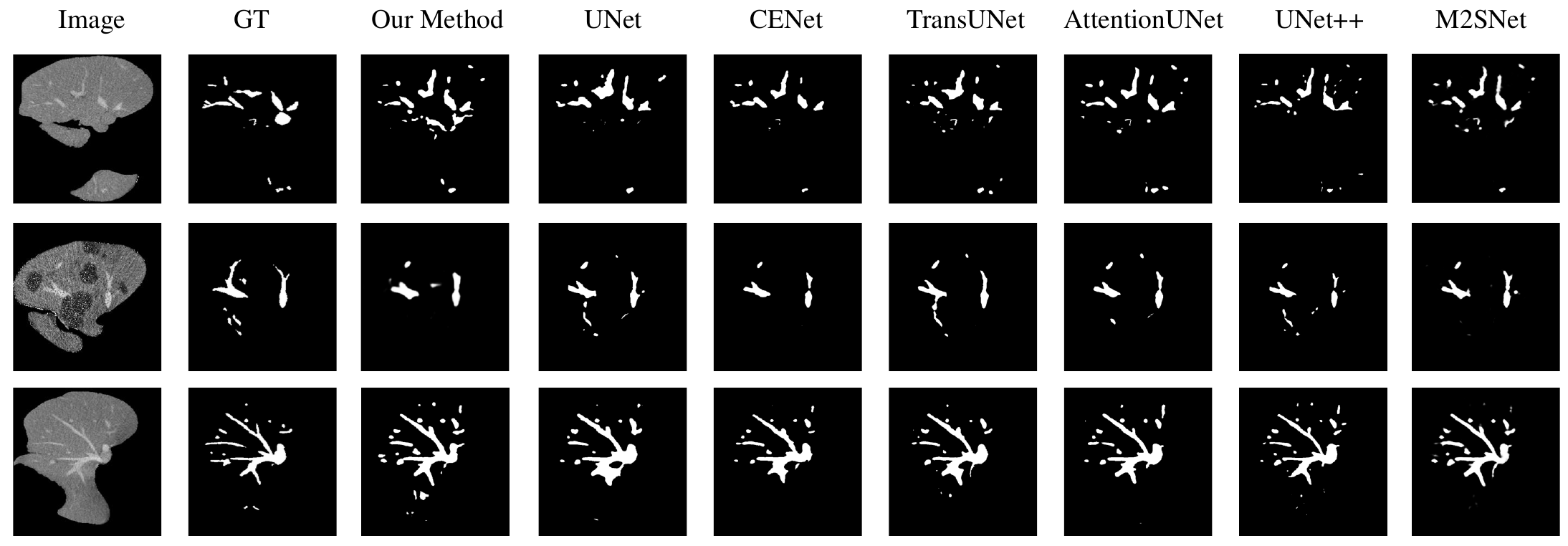}
	\caption{The visualization results of liver vessel segmentation}
	\label{fig:Figure10}
\end{figure}

For the liver vessel segmentation part of the liver cancer diagnostic assistant, this study was validated on the 3Dircadb dataset. The initial learning rate was set to 0.001, with dynamic decay, using the Adam optimizer for 400 epochs. The obtained metrics are shown in the Table~\ref{tab:liver vessel seg performance}, and the visual comparison of our method with other methods is shown in Figure~\ref{fig:Figure10}.
\begin{table}
	\centering
	\caption{Vessel segmentation performance of different models on the 3Dircadb dataset.}
	\begin{tabular}{cccccc}
		\toprule
		Model        & DICE      & IOU  & ACC  & REC      & PRE     \\
		\midrule
		UNET   & 0.516     & 0.360  & 0.986 & 0.435 & 0.744\\
		CENet   & 0.517     & 0.365  & 0.987   & 0.424    &\textbf{0.756} \\
		TransUNet & 0.537   & 0.383  & 0.986  &0.513 &0.664\\
		AttentionUNet & 0.550  & 0.392  & 0.987    &0.491 &0.719\\
		UNet++ &0.513 &0.357 &0.986 &0.449 &0.697\\
		M2SNET &0.569 &0.412 &0.987 &0.517 &0.726\\
		Our method & \textbf{0.594}    &\textbf{0.436}  &\textbf{0.988} &\textbf{0.550} &0.717\\
		\bottomrule
	\end{tabular}
	\label{tab:liver vessel seg performance}
\end{table}

\subsection{LLM results}
\begin{figure}[htb!]
	\centering
	\includegraphics[width=0.85\linewidth]{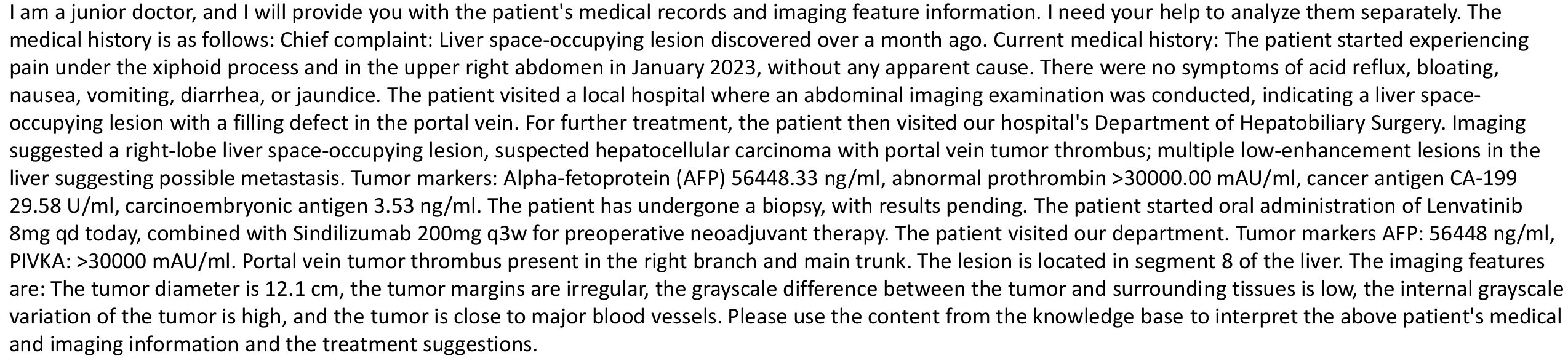}
	\caption{Qusetion}
	\label{fig:Figure11}
\end{figure}

\begin{figure}[htb!]
	\centering
	\includegraphics[width=0.85\linewidth]{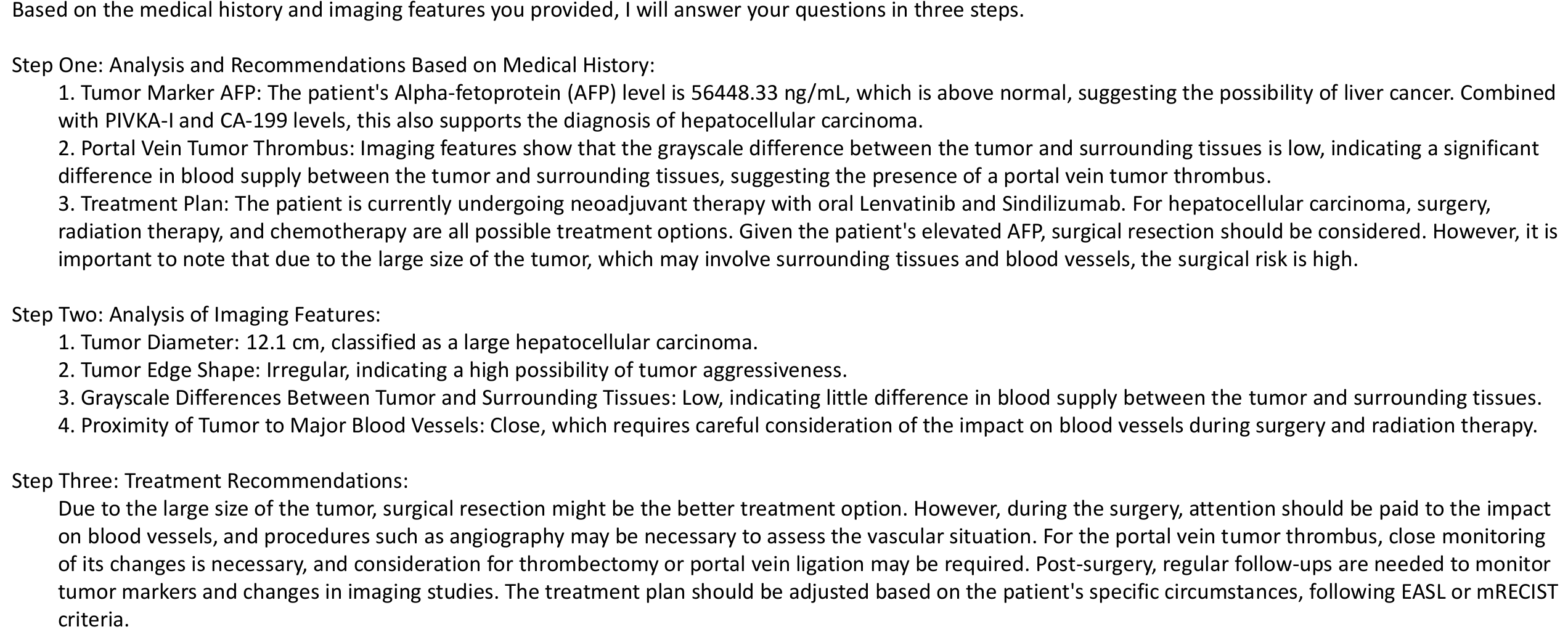}
	\caption{Our Method's Answer}
	\label{fig:Figure12}
\end{figure}

\begin{figure}[htb!]
	\centering
	\includegraphics[width=0.85\linewidth]{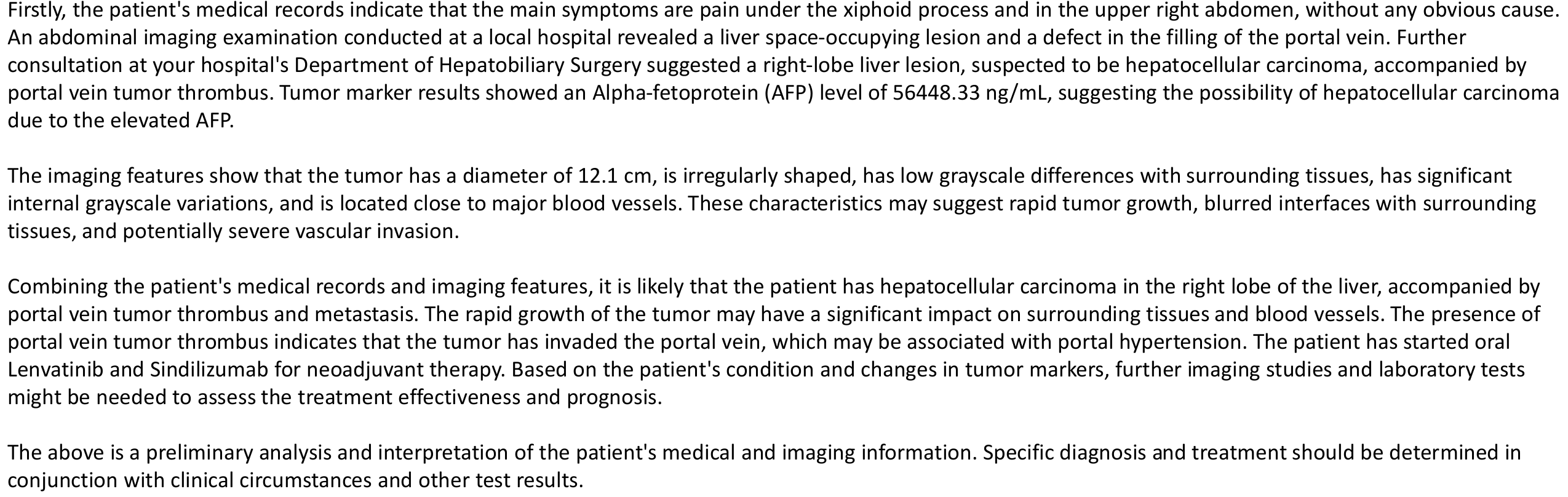}
	\caption{Control Method's Answer}
	\label{fig:Figure13}
\end{figure}

To validate the design of the large model handling part in the liver cancer diagnostic assistant, this method demonstrates two key features: one is that the large model answers based on filtered, reliable, and accurate medical knowledge rather than its existing knowledge; the other is that this method uses a chain of thought template that mimics the thinking process of experienced doctors. The question content is shown in Figure~\ref{fig:Figure11}., the original answer from the ChatGLM-6B model is shown in Figure~\ref{fig:Figure12}, and the result using ChatGLM-6B with this method is shown in Figure~\ref{fig:Figure13}. Note that since the medical knowledge and patient information referenced in this study are in Chinese, the questions and answers were also in Chinese and have been translated into English here. Besides the intuitive superiority of this method in the answers, medical experts were also asked to score the results. Two important indicators were selected: the quality of medical record and imaging information interpretation, and the rationality of the given treatment methods. After processing the information of 10 patients, questions were asked, and the results obtained from this method and the control method were shuffled and given to the senior doctor for scoring, with a maximum score of 10 points.The scoring results are shown in the table.

\begin{table}[htb!]
	\centering
	\caption{Doctor Evaluation Results}
	\begin{tabular}{|c|p{6cm}|p{6cm}|}
		\hline
		Method & \centering Average Score for Interpretation Quality of Medical and Imaging Information & \centering Average Score for Reasonableness of Treatment Plan Proposed \tabularnewline
		\hline
		This Study's Method & 7.2 & 6.5 \\
		\hline
		Control Method & 5.9 & 4.2 \\
		\hline
	\end{tabular}
	\label{tab:doctor_evaluation_results}
\end{table}
\section{DISCUSSION}
\subsection{Liver tumor segmentation}
From the quantitative metrics, the liver tumor segmentation method proposed in this study achieved better segmentation performance on both public and private datasets compared to existing networks. The data distributions of the two datasets exhibit significant differences, including variations in CT noise disturbances and liver contrast in patients. The high performance on both datasets indicates that the proposed method is highly robust and performs excellently for liver tumor segmentation tasks.

From the visualization results, the proposed segmentation method achieves more accurate edge segmentation that aligns better with the shapes in the labels. This is partly due to the use of an edge enhancement module for edge sharpening and partly due to the iterative removal of false positives and false negatives. The figures also show the absence of false positive and false negative objects that appear in other segmentation networks, further proving the effectiveness of the proposed method in filtering ambiguous pixels.

From the results of the ablation experiments, the Boundary Enhanced Module, Focusing Module, and Positioning Module all contributed to the improvement of the model performance, with the initial positioning module having the most significant impact.
The improvement in the performance of the tumor segmentation model proposed in this study can be attributed to the following points. First, the method was specifically adjusted based on the characteristics of the dataset, such as edge enhancement for blurred edges. Second, the method employed advanced attention mechanisms, which better captured channel and spatial features. Third, the iterative refinement approach after rough positioning effectively filtered ambiguous pixels.
\subsection{Liver vessel segmentation}

From the quantitative metrics of different models on the liver vessel segmentation task, it can be seen that the method proposed in this study exhibits excellent segmentation performance. It achieves the best results in DICE coefficient, IoU, accuracy, and recall, with a slight decrease in precision. Since the DICE coefficient is numerically equivalent to the  $\text{F1\_Score}$ in binary classification problems, and the  $\text{F1\_Score}$ represents the harmonic mean of recall and precision, the highest DICE coefficient in this experiment signifies the best overall performance in terms of precision and recall, rather than sacrificing precision to improve recall. Moreover, in balancing the metrics of precision and recall, improving recall in the process of detecting vessels is more beneficial for doctors. The various metrics in this experiment demonstrate the excellent capability of this method in segmenting small objects.

From the visualization results, it can be observed that, compared to other methods, this study extracts some easily overlooked vascular information well, with shapes closer to the real labels. This achieves the goal of enhancing differential information and improving segmentation performance when the network is proposed.

The reasons for the superior performance of this method are mainly due to its ability to extract more important features by utilizing information differences across different scales and levels, and the use of a weighted loss function that better focuses on small objects easily overlooked after multiple downsamplings. For future optimization, this method could consider removing some cumbersome structures. The network includes a substantial number of differences, skip connections, and calculations involving different receptive fields. It might be worth exploring whether simplifying some parts could achieve similar performance, thereby improving the training and inference speed of the model.

\subsection{LLM processing}
From the comparison of answers generated by our method and the control method, it is evident that the control method's analysis of image features is merely repetitive and lacks detailed analysis. The suggestions for treatment plans are very superficial and contain relatively little useful information.
In contrast, our method is well-organized, provides detailed analysis of medical record and imaging information, and offers high-quality answers for treatment suggestions, containing more useful information. Additionally, the references are all medical materials screened by doctors, enhancing the credibility and interpretability of the answers.
The doctors' scores also reflect significant improvements in the two key indicators,quality of medical record and imaging information interpretation, and the rationality of the given treatment plans.And it demonstrates that our method has a certain utility in aiding liver cancer diagnosis.

After being informed about the methods corresponding to the results, the senior doctor noted that the results generated by this study contained more comprehensive information and were more in line with diagnostic consensus and doctors' thinking patterns. In terms of medical record and imaging analysis, the method in this study offered a clearer description of the imaging characteristics of liver tumors, generally providing a diagnostically inclined interpretation, which has a certain advantage over the control method. Regarding treatment plans, the suggestions provided were relatively comprehensive and the treatment approach was better than that of the control method, although there is still room for improvement.

Moreover, this part only involves analyzing the information obtained by extracting image features using the framework of our method. If we solely rely on a large language model, it cannot perceive imaging information or analyze it in conjunction with medical record information.

\section{CONCLUSION}
To overcome the current limitations of large models in understanding the semantic information of medical images for specific diseases and to meet the specialized needs of doctors in this field, this study proposes a diagnostic framework in the domain of liver cancer that combines small models and large models to enhance diagnostic efficacy.

At first, to more accurately obtain a personalized knowledge base for patients and enhance the perception of semantic information in their CT images, the study incorporates small models for more precise extraction. The liver tumor segmentation method in this study shows improvement in addressing the problem of edge blurriness and missegmentation in liver tumor segmentation. Similarly, the liver vessel segmentation method improves the extraction of fine vessel information, which is often challenging. After obtaining the segmentation results, quantitative formulas are used to mimic the perception of experienced human doctors for feature extraction from the segmentation results, which, combined with medical records, forms the patient's personalized knowledge base.

Finally, large models are used to analyze patients’ personalized information. To ensure the reliability of the reference materials for the large model, RAG (Retrieval-Augmented Generation) technology is employed to answer questions based on trusted diagnostic knowledge and cases selected by doctors. To improve the quality of the large model's responses, specific prompts are designed to mimic the answers of experienced doctors. This process references three types of knowledge bases: the patient's personalized knowledge base, general diagnostic knowledge in the field, and private diagnostic case knowledge from experienced doctors.

In comparison, this study's Diagnosis Assistant for Liver Cancer is particularly beneficial for assisting less experienced doctors in diagnosing liver cancer in regions with limited medical resources.

\clearpage

\begin{thebibliography}{37}
\providecommand{\natexlab}[1]{#1}
\providecommand{\url}[1]{\texttt{#1}}
\expandafter\ifx\csname urlstyle\endcsname\relax
  \providecommand{\doi}[1]{doi: #1}\else
  \providecommand{\doi}{doi: \begingroup \urlstyle{rm}\Url}\fi

\bibitem[Anter et~al.(2013)Anter, Azar, Hassanien, El-Bendary, and ElSoud]{anter2013automatic}
A.~M. Anter, A.~T. Azar, A.~E. Hassanien, N.~El-Bendary, and M.~A. ElSoud.
\newblock Automatic computer aided segmentation for liver and hepatic lesions using hybrid segmentations techniques.
\newblock In \emph{2013 Federated Conference on Computer Science and Information Systems}, pages 193--198. IEEE, 2013.

\bibitem[Bao et~al.(2023)Bao, Chen, Xiao, Ren, Wu, Zhong, Peng, Huang, and Wei]{bao2023disc}
Z.~Bao, W.~Chen, S.~Xiao, K.~Ren, J.~Wu, C.~Zhong, J.~Peng, X.~Huang, and Z.~Wei.
\newblock Disc-medllm: Bridging general large language models and real-world medical consultation.
\newblock \emph{arXiv preprint arXiv:2308.14346}, 2023.

\bibitem[Brown et~al.(2020)Brown, Mann, Ryder, Subbiah, Kaplan, Dhariwal, Neelakantan, Shyam, Sastry, Askell, et~al.]{brown2020language}
T.~Brown, B.~Mann, N.~Ryder, M.~Subbiah, J.~D. Kaplan, P.~Dhariwal, A.~Neelakantan, P.~Shyam, G.~Sastry, A.~Askell, et~al.
\newblock Language models are few-shot learners.
\newblock \emph{Advances in neural information processing systems}, 33:\penalty0 1877--1901, 2020.

\bibitem[Budak et~al.(2020)Budak, Guo, Tanyildizi, and {\c{S}}eng{\"u}r]{budak2020cascaded}
{\"U}.~Budak, Y.~Guo, E.~Tanyildizi, and A.~{\c{S}}eng{\"u}r.
\newblock Cascaded deep convolutional encoder-decoder neural networks for efficient liver tumor segmentation.
\newblock \emph{Medical Hypotheses}, 134:\penalty0 109431, 2020.

\bibitem[Chen et~al.(2021)Chen, Lu, Yu, Luo, Adeli, Wang, Lu, Yuille, and Zhou]{chen2021transunet}
J.~Chen, Y.~Lu, Q.~Yu, X.~Luo, E.~Adeli, Y.~Wang, L.~Lu, A.~L. Yuille, and Y.~Zhou.
\newblock Transunet: Transformers make strong encoders for medical image segmentation.
\newblock \emph{arXiv preprint arXiv:2102.04306}, 2021.

\bibitem[Chen et~al.(2023)Chen, Wang, Gao, Jiang, Chen, Zhang, Song, Xie, Kong, Li, et~al.]{chen2023huatuogpt}
J.~Chen, X.~Wang, A.~Gao, F.~Jiang, S.~Chen, H.~Zhang, D.~Song, W.~Xie, C.~Kong, J.~Li, et~al.
\newblock Huatuogpt-ii, one-stage training for medical adaption of llms.
\newblock \emph{arXiv preprint arXiv:2311.09774}, 2023.

\bibitem[Choudhary et~al.(2008)Choudhary, Moretto, Ferrarese, and Zamboni]{choudhary2008entropy}
A.~Choudhary, N.~Moretto, F.~P. Ferrarese, and G.~A. Zamboni.
\newblock An entropy based multi-thresholding method for semi-automatic segmentation of liver tumors.
\newblock In \emph{MICCAI Workshop}, volume~41, pages 43--49, 2008.

\bibitem[Foruzan et~al.(2012)Foruzan, Zoroofi, Sato, and Hori]{foruzan2012hessian}
A.~H. Foruzan, R.~A. Zoroofi, Y.~Sato, and M.~Hori.
\newblock A hessian-based filter for vascular segmentation of noisy hepatic ct scans.
\newblock \emph{International Journal of Computer Assisted Radiology and Surgery}, 7:\penalty0 199--205, 2012.

\bibitem[Friman et~al.(2010)Friman, Hindennach, K{\"u}hnel, and Peitgen]{friman2010multiple}
O.~Friman, M.~Hindennach, C.~K{\"u}hnel, and H.-O. Peitgen.
\newblock Multiple hypothesis template tracking of small 3d vessel structures.
\newblock \emph{Medical Image Analysis}, 14\penalty0 (2):\penalty0 160--171, 2010.

\bibitem[Guo et~al.(1998)Guo, Chen, Lee, and Tsai]{guo19983}
J.-K. Guo, C.-H. Chen, J.-D. Lee, and J.-M. Tsai.
\newblock 3-d image reconstruction of brain blood vessels from angiograms.
\newblock \emph{Computers \& Mathematics with Applications}, 35\penalty0 (8):\penalty0 79--94, 1998.

\bibitem[Guo et~al.(2020)Guo, Xiao, Zhang, Chen, Wang, and Wang]{guo2020novel}
X.~Guo, R.~Xiao, T.~Zhang, C.~Chen, J.~Wang, and Z.~Wang.
\newblock A novel method to model hepatic vascular network using vessel segmentation, thinning, and completion.
\newblock \emph{Medical \& Biological Engineering \& Computing}, 58:\penalty0 709--724, 2020.

\bibitem[Kitrungrotsakul et~al.(2017)Kitrungrotsakul, Han, Iwamoto, Foruzan, Lin, and Chen]{kitrungrotsakul2017robust}
T.~Kitrungrotsakul, X.-H. Han, Y.~Iwamoto, A.~H. Foruzan, L.~Lin, and Y.-W. Chen.
\newblock Robust hepatic vessel segmentation using multi deep convolution network.
\newblock In \emph{Medical Imaging 2017: Biomedical Applications in Molecular, Structural, and Functional Imaging}, volume 10137, pages 269--274. SPIE, 2017.

\bibitem[Kushnure and Talbar(2021)]{kushnure2021ms}
D.~T. Kushnure and S.~N. Talbar.
\newblock Ms-unet: A multi-scale unet with feature recalibration approach for automatic liver and tumor segmentation in ct images.
\newblock \emph{Computerized Medical Imaging and Graphics}, 89:\penalty0 101885, 2021.

\bibitem[Lei et~al.(2021)Lei, Wang, Zhang, Wan, Liu, and Nandi]{lei2021defed}
T.~Lei, R.~Wang, Y.~Zhang, Y.~Wan, C.~Liu, and A.~K. Nandi.
\newblock Defed-net: Deformable encoder-decoder network for liver and liver tumor segmentation.
\newblock \emph{IEEE Transactions on Radiation and Plasma Medical Sciences}, 6\penalty0 (1):\penalty0 68--78, 2021.

\bibitem[Li et~al.(2012)Li, Chui, Chang, and Ong]{li2012new}
B.~N. Li, C.~K. Chui, S.~Chang, and S.~H. Ong.
\newblock A new unified level set method for semi-automatic liver tumor segmentation on contrast-enhanced ct images.
\newblock \emph{Expert Systems with Applications}, 39\penalty0 (10):\penalty0 9661--9668, 2012.

\bibitem[Li et~al.(2020)Li, Tan, Chen, Luo, Gao, Jia, and Wang]{li2020attention}
C.~Li, Y.~Tan, W.~Chen, X.~Luo, Y.~Gao, X.~Jia, and Z.~Wang.
\newblock Attention unet++: A nested attention-aware u-net for liver ct image segmentation.
\newblock In \emph{2020 IEEE International Conference on Image Processing (ICIP)}, pages 345--349. IEEE, 2020.

\bibitem[Li et~al.(2019)Li, Zhang, and Gao]{li2019vessel}
J.~Li, M.~Zhang, and Y.~Gao.
\newblock Vessel segmentation of liver ct images by hessian-based enhancement.
\newblock In \emph{Image and Graphics: 10th International Conference, ICIG 2019, Beijing, China, August 23--25, 2019, Proceedings, Part III 10}, pages 442--455. Springer, 2019.

\bibitem[Li et~al.(2017)Li, Jiang, and Pang]{li2017joint}
S.~Li, H.~Jiang, and W.~Pang.
\newblock Joint multiple fully connected convolutional neural network with extreme learning machine for hepatocellular carcinoma nuclei grading.
\newblock \emph{Computers in Biology and Medicine}, 84:\penalty0 156--167, 2017.

\bibitem[Liu et~al.(2021)Liu, Shao, and Hoffmann]{liu2021global}
Y.~Liu, Z.~Shao, and N.~Hoffmann.
\newblock Global attention mechanism: Retain information to enhance channel-spatial interactions.
\newblock \emph{arXiv preprint arXiv:2112.05561}, 2021.

\bibitem[Mei et~al.(2021)Mei, Ji, Wei, Yang, Wei, and Fan]{mei2021camouflaged}
H.~Mei, G.-P. Ji, Z.~Wei, X.~Yang, X.~Wei, and D.-P. Fan.
\newblock Camouflaged object segmentation with distraction mining.
\newblock In \emph{Proceedings of the IEEE/CVF Conference on Computer Vision and Pattern Recognition}, pages 8772--8781, 2021.

\bibitem[Moghe et~al.(2011)Moghe, Singhai, and Shrivastava]{moghe2011automatic}
A.~A. Moghe, J.~Singhai, and S.~Shrivastava.
\newblock Automatic threshold based liver lesion segmentation in abdominal 2d-ct images.
\newblock \emph{International Journal of Image Processing (IJIP)}, 5\penalty0 (2):\penalty0 166, 2011.

\bibitem[Oliveira et~al.(2011)Oliveira, Feitosa, and Correia]{oliveira2011segmentation}
D.~A. Oliveira, R.~Q. Feitosa, and M.~M. Correia.
\newblock Segmentation of liver, its vessels and lesions from ct images for surgical planning.
\newblock \emph{Biomedical Engineering Online}, 10:\penalty0 1--23, 2011.

\bibitem[Shimizu et~al.(2008)Shimizu, Narihira, Furukawa, Kobatake, Nawano, and Shinozaki]{shimizu2008ensemble}
A.~Shimizu, T.~Narihira, D.~Furukawa, H.~Kobatake, S.~Nawano, and K.~Shinozaki.
\newblock Ensemble segmentation using adaboost with application to liver lesion extraction from a ct volume.
\newblock In \emph{Proc. MICCAI Workshop on 3D Segmentation in the Clinic: A Grand Challenge II., NY, USA}, 2008.

\bibitem[Singhal et~al.(2022)Singhal, Azizi, Tu, Mahdavi, Wei, Chung, Scales, Tanwani, Cole-Lewis, Pfohl, et~al.]{singhal2022large}
K.~Singhal, S.~Azizi, T.~Tu, S.~S. Mahdavi, J.~Wei, H.~W. Chung, N.~Scales, A.~Tanwani, H.~Cole-Lewis, S.~Pfohl, et~al.
\newblock Large language models encode clinical knowledge.
\newblock \emph{arXiv preprint arXiv:2212.13138}, 2022.

\bibitem[Sun et~al.(2017)Sun, Guo, Zhang, Li, Ma, and Li]{sun2017liver}
C.~Sun, S.~Guo, H.~Zhang, J.~Li, S.~Ma, and X.~Li.
\newblock Liver lesion segmentation in ct images with mk-fcn.
\newblock In \emph{2017 IEEE 2nd Advanced Information Technology, Electronic and Automation Control Conference (IAEAC)}, pages 1794--1798. IEEE, 2017.

\bibitem[Tian et~al.(2023)Tian, Gan, Song, Zhang, and Zhang]{tian2023chimed}
Y.~Tian, R.~Gan, Y.~Song, J.~Zhang, and Y.~Zhang.
\newblock Chimed-gpt: A chinese medical large language model with full training regime and better alignment to human preferences.
\newblock \emph{arXiv preprint arXiv:2311.06025}, 2023.

\bibitem[Valanarasu et~al.(2021)Valanarasu, Oza, Hacihaliloglu, and Patel]{valanarasu2021medical}
J.~M.~J. Valanarasu, P.~Oza, I.~Hacihaliloglu, and V.~M. Patel.
\newblock Medical transformer: Gated axial-attention for medical image segmentation.
\newblock In \emph{Medical Image Computing and Computer Assisted Intervention--MICCAI 2021: 24th International Conference, Strasbourg, France, September 27--October 1, 2021, Proceedings, Part I 24}, pages 36--46. Springer, 2021.

\bibitem[Wang et~al.(2023)Wang, Zhao, Ouyang, Wang, and Shen]{wang2023chatcad}
S.~Wang, Z.~Zhao, X.~Ouyang, Q.~Wang, and D.~Shen.
\newblock Chatcad: Interactive computer-aided diagnosis on medical image using large language models.
\newblock \emph{arXiv preprint arXiv:2302.07257}, 2023.

\bibitem[Wilson and Noble(1997)]{wilson1997segmentation}
D.~L. Wilson and J.~A. Noble.
\newblock Segmentation of cerebral vessels and aneurysms from mr angiography data.
\newblock In \emph{Information Processing in Medical Imaging: 15th International Conference, IPMI'97 Poultney, Vermont, USA, June 9--13, 1997 Proceedings 15}, pages 423--428. Springer, 1997.

\bibitem[Xu et~al.(2020)Xu, Wang, Chi, and Hua]{xu2020training}
M.~Xu, Y.~Wang, Y.~Chi, and X.~Hua.
\newblock Training liver vessel segmentation deep neural networks on noisy labels from contrast ct imaging.
\newblock In \emph{2020 IEEE 17th International Symposium on Biomedical Imaging (ISBI)}, pages 1552--1555. IEEE, 2020.

\bibitem[Xu et~al.(2021)Xu, Lu, Wang, Luo, Jayender, Ma, Zheng, and Li]{xu2021noisy}
Z.~Xu, D.~Lu, Y.~Wang, J.~Luo, J.~Jayender, K.~Ma, Y.~Zheng, and X.~Li.
\newblock Noisy labels are treasure: mean-teacher-assisted confident learning for hepatic vessel segmentation.
\newblock In \emph{Medical Image Computing and Computer Assisted Intervention--MICCAI 2021: 24th International Conference, Strasbourg, France, September 27--October 1, 2021, Proceedings, Part I 24}, pages 3--13. Springer, 2021.

\bibitem[Yan et~al.(2020)Yan, Wang, Zhang, Luo, Xu, Xu, Zhang, Shi, Zhang, and You]{yan2020attention}
Q.~Yan, B.~Wang, W.~Zhang, C.~Luo, W.~Xu, Z.~Xu, Y.~Zhang, Q.~Shi, L.~Zhang, and Z.~You.
\newblock Attention-guided deep neural network with multi-scale feature fusion for liver vessel segmentation.
\newblock \emph{IEEE Journal of Biomedical and Health Informatics}, 25\penalty0 (7):\penalty0 2629--2642, 2020.

\bibitem[Yu et~al.(2019)Yu, Fang, Liu, Gao, Zheng, and Wang]{yu2019liver}
W.~Yu, B.~Fang, Y.~Liu, M.~Gao, S.~Zheng, and Y.~Wang.
\newblock Liver vessels segmentation based on 3d residual u-net.
\newblock In \emph{2019 IEEE International Conference on Image Processing (ICIP)}, pages 250--254. IEEE, 2019.

\bibitem[Yuan(2017)]{yuan2017hierarchical}
Y.~Yuan.
\newblock Hierarchical convolutional-deconvolutional neural networks for automatic liver and tumor segmentation.
\newblock \emph{arXiv preprint arXiv:1710.04540}, 2017.

\bibitem[Zeng et~al.(2022)Zeng, Liu, Du, Wang, Lai, Ding, Yang, Xu, Zheng, Xia, et~al.]{zeng2022glm}
A.~Zeng, X.~Liu, Z.~Du, Z.~Wang, H.~Lai, M.~Ding, Z.~Yang, Y.~Xu, W.~Zheng, X.~Xia, et~al.
\newblock Glm-130b: An open bilingual pre-trained model.
\newblock \emph{arXiv preprint arXiv:2210.02414}, 2022.

\bibitem[Zeng et~al.(2018)Zeng, Liao, Tang, Zhao, Liao, Chen, and Liang]{zeng2018automatic}
Y.-z. Zeng, S.-h. Liao, P.~Tang, Y.-q. Zhao, M.~Liao, Y.~Chen, and Y.-x. Liang.
\newblock Automatic liver vessel segmentation using 3d region growing and hybrid active contour model.
\newblock \emph{Computers in Biology and Medicine}, 97:\penalty0 63--73, 2018.

\bibitem[Zhou et~al.(2018)Zhou, Rahman~Siddiquee, Tajbakhsh, and Liang]{zhou2018unet++}
Z.~Zhou, M.~M. Rahman~Siddiquee, N.~Tajbakhsh, and J.~Liang.
\newblock Unet++: A nested u-net architecture for medical image segmentation.
\newblock In \emph{Deep Learning in Medical Image Analysis and Multimodal Learning for Clinical Decision Support: 4th International Workshop, DLMIA 2018, and 8th International Workshop, ML-CDS 2018, Held in Conjunction with MICCAI 2018, Granada, Spain, September 20, 2018, Proceedings 4}, pages 3--11. Springer, 2018.

\end{thebibliography}




\end{document}